\documentclass[twocolumn,amsmath,amssymb,prb,showpacs,preprintnumbers,aps]{revtex4}
\usepackage{graphicx}
\usepackage{bm}
\usepackage{amsmath,amssymb} 
\usepackage{dcolumn}

\begin{document}

\title{Vibrational modes of metal nanoshells and bimetallic core-shell nanoparticles} 

\author{Arman S. Kirakosyan$^{a,b}$, Tigran V. Shahbazyan$^a$}

\affiliation{$^a$Department of Physics, Jackson State University, Jackson, MS
39217, USA\\
$^b$Department of Physics, Yerevan State University, 1 Alex Manoogian
St., Yerevan, 375025, Armenia}  

%
\begin{abstract}
We study theoretically spectrum of radial vibrational modes in composite metal nanostructures such as bimetallic core-shell particles and metal nanoshells with dielectric core in an environment. We calculate frequencies and damping rates of fundamental (breathing) modes for these nanostructures along with those of two higher-order modes. For metal nanoshells, we find that the breathing mode frequency is always lower than the one for solid particles of the same size, while the damping is higher and increases with reduction of the shell thickness.  We identify  two regimes that can be characterized as weakly damped and overdamped vibrations in the presence of external medium. For bimetalllic particles, we find periodic dependence of frequency and damping rate on  the shell thickness with period  determined by mode number. For both types of nanostructures, the frequency of higher modes is nearly independent of the environment, while the damping rate shows strong sensitivity to outside medium.
\end{abstract}
\maketitle

\section{Introduction}
\label{intro}

Composite  nanostructures comprised of alternating metal and dielectric constituents such as, e.g., bimetallic spherical particles, metal nanoshells,\cite{halas-prl97}  nanorice,\cite{halas-nanorice}, and structures with more than two constituents (nano-matryoshka)\cite{halas-matryoshka}, are among the highlights of nanomaterials with versatile optical and mechanical properties. Similar to solid metal nanoparticles, the absorption and scattering of light by composite nanostructures are dominated by the surface plasmon resonance (SPR). These structures offer new possibilities for various applications due to tunability  of their optical properties which can be controlled during the manufacturing process by varying constituent materials or geometry.\cite{halas-prb02,prodan-nl02} For example, by changing  shell thickness, the SPR position in nanoshells and bimetallic particles can be changed in wide frequency range. Recent measurements on single nanoshells\cite{klar-nl04,halas-nl04} indicated enhanced sensitivity of their scattering spectra to the environment, while the SPR redshift resulted in a narrowing of their resonance lineshape as compared to  solid gold particles.\cite{klar-prl98} These unique optical characteristics of composite structures spurred numerous proposals for various applications, e.g., in optomechanics,\cite{halas-ap01} sensing,\cite{sun-ac02,halas-apl03} drug delivery\cite{halas-ac03,halas-nl02} and phototherapy.\cite{west-pnas03}

While the optical response of core-shell nanostructures has been extensively studied, much less is known about their acoustical properties.\cite{vallee-jcp99,hartland-jcp99,vallee-jcp01} In fact, the low frequency vibrational modes of composite nanoparticles bear a unique signature of their structural and mechanical characteristics directly reflecting the impact of confinement on the ionic movement.\cite{vallee-jcp01,elsayed-nl04,orrit-prl05} This is in contrast to the optical response which is mainly determined by electronic excitations. Thus the vibrational modes constitute an additional source of information that is particularly important for hybrid or layered systems with structurally distinct constituents, such as, e.g., bimetallic nanoparticles and nanoshells. Recent pump-probe spectroscopy measurements of fundamental (breathing) mode frequency and damping in bimetallic particles \cite{hartland-bimetal1,hartland-bimetal2,hartland-nl07} and gold nanoshells \cite{guillon-nl07} revealed significantly different, from solid nanoparticles, acoustical signature that could be used as a tool complimentary to optical methods.

In this paper we study the spectrum of radial vibrational modes of core-shell nanoparticles, including metal nanoshells with dielectric core and bimetallic spherical particles, embedded in an environment. Acoustic vibrational modes in nanoparticles can be impulsively launched by a laser pulse and monitored using the pump-probe spectroscopic technique.\cite{vallee-jcp99,hartland-jcp99,vallee-jcp01,elsayed-nl04,hartland-bimetal1,hartland-bimetal2} After initial stage of rapid expansion, the nanoparticle  undergoes radial oscillations around its new equilibrium size. The periodic change in volume translates into a modulation, in time, of the SPR energy and manifests itself in the coherent oscillations of differential transmission.\cite{vallee-jpcb01} Since  the laser spot is usually much larger than the nanoparticle size, the initial expansion is homogeneous, so that predominantly the fundamental ($n=0$) breathing mode, corresponding to oscillations of the nanoparticle volume as a whole, is excited. In solid nanoparticles, frequencies of radial modes are inversely proportional to the nanoparticle radius, $ \omega_{n} \approx \pi (n+1)c_{L}/R$,\cite{love-elast} where $ c_{L} $ is longitudinal sound velocity. When a nanoparticle is embedded in dielectric medium, the oscillations are damped, due to the transfer of latice energy to the environment, with damping rate\cite{dubrovskiy81,vallee-pb02} $ \gamma \propto R^{-1} $. For homogeneous intial expansion, the higher vibrational modes are much less pronounced. With the pump-probe technique,  the $ n=1 $ and $ n=2 $ modes were recently detected in Ag spherical particles, while the higher radial modes (up to $n=4$) were observed using Raman scattering technique.\cite{vallee-ass04}

In this paper we present a detailed analysis of energy spectrum for the first three ($ n=0,1,2 $) lowest  radial vibrational modes in metal nanoshells and bimetallic core-shell nanoparticles in different environments. We find that the modes eigen-energies exhibit a strong dependence on the aspect ratio, $\kappa=R_1/R_2$, where $R_1$ and $R_2$ are the inner and outer radii, respectively. For nanoshells,  the fundamental mode energy is considerably lower than that for solid particles, while the damping rate exhibits a more complicated behavior; at $ \kappa \approx 0.4 $ it has a minimum, while for larger $\kappa$, the damping rate sharply increases,  $ \gamma \propto d^{-1} $, where $d$ is the shell thickness.  For thin shells, we find two regimes, weakly damped and overdamped, characterized by the nanoshell mass relative to displaced mass of the outer medium. These regimes are absent for higher modes, $ n \geq 1 $, where both frequency and damping rate are inversely proportional to the shell thickness.  At the same time, the impact of environment on higher modes frequencies is negligible, while the damping rates are strongly affected by outside dielectric.  For thin shells, the fundamental mode carries most of the oscillator strength, implying enhanced oscillations amplitude as compared to solid particles. However, the higher modes have relatively larger oscillator strengths and thus should be better detected. 

For bimetallic Au/Ag nanoparticles, we find that, in a wide range of aspect ratios, the fundamental mode frequency and damping rate are higher than those for either Au or Ag solid particles. For higher modes, both frequency and damping exhibit periodic oscillations with aspect ratio, in contrast to nanoshells, with period determined by the mode number $ n $. Similar to nanoshells, for higher modes, the environment significantly affects only the damping rate. The oscillator strengths also exhibit periodic dependence on $\kappa$ and have maximum values for thin shells.

The paper is organized as follows. In section \ref{sec:spectrum},  equations determining the complex eigenvalues for radial vibrational modes of spherical core-shell nanoparticles and expressions for their oscillator strength are derived.  The analysis of radial modes for Au nanoshells with dielectric core is presented in section \ref{sec:nanoshell}. In section \ref{sec:bimetallic},  the results for bimetallic Au/Ag core-shell nanoparticles are discussed. Section \ref{sec:conclusions} concludes the paper.

\section{Spectrum of radial vibrational modes of a core-shell nanoparticle}
\label{sec:spectrum}
We consider radial normal modes of a spherical core-shell particle with core extending up to inner radius $R_1$ and shell up to outer radius $R_2$ embedded in a dielectric medium. At zero angular momentum ($l=0$), the motion of nanoparticle boundaries is described by the radial displacement $u(r)=\varphi^{\prime}(r)$ (prime stands for derivative),  where displacement potential $\varphi$ satisfies the Helmholtz equation,
\begin{equation}
\varphi'' + \frac{2\varphi'}{r} + k^{2} \varphi = 0,
\label{Helm}
\end{equation}
where $k^{(i)} = \omega/c_L^{(i)}$ is the wave-vector and longitudinal sound velocity $c_L^{(i)}$ in $i$th medium. The boundary conditions demand continuity of both the displacement $u(r)$ and the radial diagonal component of the stress tensor,
\begin{equation}
\sigma_{rr} = \rho \biggl[c_L^2 u' + (c_L^2 - 2 c_T^2)\, \frac{2u}{r}
\biggr],
\label{stress}
\end{equation}
across the core/shell and shell/medium interfaces. Here $\rho$ and $c_{L,T}$ are the corresponding density and longitudinal/transverse sound velocities. In the core, shell, and medium regions, solutions of Eq.~(\ref{Helm}) are, respectively, of the form 
\begin{equation}
\label{solutions}
\varphi \sim \frac{\sin k^{(c)}r}{r},~
\varphi\sim\frac{\sin\left(k^{(s)}r+\phi\right)}{r},~  
\varphi \sim \frac{e^{ik^{(m)}r}}{r},
\end{equation}
where $\phi$ is phase shift.  Matching $u=\varphi^{\prime}$ and $\sigma_{rr}$ at the inner radius $R_1$ and outer radius $R_2$ yields equations for eigenvalues 
\begin{eqnarray}
\frac{\xi^2 \kappa^2}{\xi \kappa \cot(\xi \kappa + \phi) - 1}
-
\frac{\eta_{c}\xi^2 \kappa^2}{(\xi \kappa/\alpha_{c})
\cot(\xi  \kappa/\alpha_c) - 1} +\chi_c =0,
\nonumber \\
\frac{\xi^2}{\xi \cot(\xi + \phi) - 1}
+
\frac{\eta_{m}\xi^2}{1 + i \xi/\alpha_m} + \chi_m = 0,
\qquad \qquad
\label{breathmodes}
\end{eqnarray}
where $\xi=k^{(s)}R_2=\omega R_2/c_L^{(s)}$ and $\kappa  = R_1/R_2$ are shorthand notations for the normalized eigenfrequencies and aspect ratio, respectively. The parameters
\begin{eqnarray}
&&
\alpha_{i} = c_{L}^{(i)}/c_{L}^{(s)},
~~
\eta_{i} = \rho^{(i)}/\rho^{(s)},
~~
\chi_{i} = 4\beta_s^2 (1- \eta_i \delta_i^2)
\nonumber \\
&&
\beta_{i}=c_{T}^{(i)}/c_{L}^{(i)},
~~
\delta_{i}=c_{T}^{(i)}/c_{T}^{(s)},
\label{parameters}
\end{eqnarray}
characterize the metal/dielectric interfaces ($i=c,s,m$ stand for
core, shell, and outer medium). The parameter $ \chi_{m} = 4 \beta_{s}^{2}(1-\mu_{m}/\mu_{s})$ can be expressed via the ratio of shear modules of outside dielectric $ \mu_{m} $ and metal shell $ \mu_{s} $, where shear module is defined as $ \mu = \rho c_{T}^{2} $. Equations (\ref{breathmodes}) determine complex eigenvalues $\xi_{n}=\xi'_{n}+i\xi''_{n}$ of composite core-shell particle, $n=0,1,\ldots$ being mode numbers, where the real and imaginary parts describe mode frequency and damping rate. 

From Eqs.~(\ref{breathmodes}), the known results for \emph{solid} particles can be easily recovered by setting $\phi=0$. In the case when density of outside medium is much smaller than that of particle, the approximate solution for eigenvalues reads
\begin{eqnarray}
\label{final-xi}
&&
\xi'_{n}\approx \pi (n+1)-\frac{4 \beta_{s}^{2}}{\pi(n+1)} + \frac{4 \beta_{s}^{2} }{\pi(n+1)}\frac{\mu_{m}}{\mu_{s}},
\\ && 
\xi''_{n}\approx \frac{\rho_{m} c_{L}^{(m)}}{\rho_{s} c_{L}^{(s)}}=\frac{Z_{m}}{Z_{s}},
\end{eqnarray}
where $ Z_{i} = \rho_{i} c_{L}^{(i)}$ is the acoustic impedance of the $i$th material. Note that for solid nanoparticles, both frequency $\omega = \xi' c_{L}/R$ and damping rate $\gamma = \xi'' c_{L}/R$ decrease with particle size and that the latter is nearly independent of mode number $ n $. 

Let us now turn to relative contributions of different eigenmodes in the total acoustical response. We assume that the laser spot is larger than the nanoparticle size and, therefore, the initial expansion is isotropic. Even so, as the nanoparticle vibrates with respect to its new equilibrium size, all radial modes are, in principle, excited. A precise determination of modes contibutions is somewhat complicated due to the fact that, for composite heterogeneous nanostructures, these modes are not necessarily orthonormal. However, since the dominant contribution comes from the fundamental (breathing) mode, the orthonormal set can be constructed perturbatively. Namely, the higher modes acquire an admixture of the fundamental one, while the latter, in this approximation, remains nearly unperturbed. The orthonormal displacements set has the form
\begin{eqnarray}
\label{newset1}
&&
v_{0}=a_{0}u_{0},
~~~~
v_{n}=a_{n}^{(0)} u_{0}+a_{n}^{(n)}u_{n},
\end{eqnarray}
where  the coefficients $a $ are given by
\begin{eqnarray}
\label{coeffs}
a_{0}=\frac{1}{\sqrt{\langle u_{0}^{2}\rangle}},~~
a_{n}^{(0)} = \frac{\nu_{n}}{\sqrt{\langle u_{0}^{2}\rangle}},
~~a_{n}^{(n)} = -\frac{\nu_{n} \sqrt{\langle u_{0}^{2}\rangle}}{\langle u_{0} u_{n}\rangle},
\end{eqnarray}
with $ \langle f \rangle = V^{-1} \int f(r) dV$, and the parameter
\begin{equation}
\nu_{n}=\left (\frac{\langle u_{0}^{2}\rangle \langle u_{n}^{2}\rangle}{\langle u_{0} u_{n}\rangle^{2}} - 1\right )^{-1/2}
\end{equation}
characterizes the admixture. The radial displacement $\delta r(t)$ (relative to average radius) can be expanded over  $v_n(r)$ as $\delta r(t)=\sum_n b_nv_n(r)e^{-i\omega_nt}$, where coefficients $b_n$ are determined from the initial condition. For homogeneous initial expansion, $\delta r(0)\propto r$, the \emph{oscillator strengths}, 
\begin{equation}
C_{n}= \frac{b_n}{\left(\sum\limits_n b_n^2\right)^{1/2}}= 
 \frac{\langle r v_{n} \rangle}{\sqrt{\langle r^2\rangle}},
\label{oscill}
\end{equation}
determine relative contributions of the corresponding modes in the total acoustical response. In the following sections, we present detailed analysis of the vibrational modes spectrum for metal nanoshells and bimetallic particles.

\section{Vibrational modes of metal nanoshell}
\label{sec:nanoshell}

The approach developed in previous section describes vibrational eigenmodes of a nanoparticle in an \emph{equilibrium} state. However, acoustical modes are usually launched by ultrashort laser pulse that results in rapid initial expansion of the particle followed by oscillations around its new equilibrium position.  In the case of a metal nanoshell with a dielectric core, the latter plays negligible role and the eigenmodes are determined mainly by the metal shell and outside medium. Indeed, the dielectric core is not directly affected by the laser pulse, but experiences thermal expansion as a result of heat transfer from the metal shell. At the same time, this expansion is much weaker than that of the metal, so that when new equilibrium size is established, the core is almost completely disengaged from the shell. This should be contrasted to bimetallic particles where the core does remain engaged after the expansion and thus contributes to the acoustical vibration spectrum (see the next section). 

Consider first the fundamental ($n=0$) mode. For the ideal case of a nanoshell in vacuum, frequency is obtained from Eq.~(\ref{breathmodes}) with stress-free inner boundary, i.e., by setting $\eta_m=\eta_c=0$ and $\chi_c=\chi_m=4\beta_s^2$. In the thin shell limit, $1-\kappa = d/R_2\ll 1$, we then recover the
known result \cite{love-elast}
\begin{equation}
\label{thin}
\xi_0= 2 \beta_s \sqrt{3 - 4\beta_s^2}.
\end{equation}
For a nanoshell in a dielectric medium, the eigenvalues are complex due to the energy exchange with the environment. For thin nanoshells, $1-\kappa = d/R_2\ll 1$, Eqs. (\ref{breathmodes}) (with $\eta_c=0$) reduces to
\begin{equation}
\frac{\chi_c}{1-\kappa}
\biggl(
\chi_m -\chi_c + \frac{\alpha_m \eta_m \xi^2}{\alpha_m + i \xi}
\biggr)
=
\biggl(
\chi_m + \frac{\alpha_m \eta_m \xi^2}{\alpha_m + i \xi}
\biggr) \xi_0^2
- \chi_c \xi^2.
\label{simple}
\end{equation}
In the typical case when the metal shell density is much higher than that of the surrounding medium, i.e., $\eta_m=\rho^{(m)}/\rho^{(s)}\ll 1$, Eq.\ (\ref{simple}) further simplifies to
\begin{equation}
x^2 - 1  =
\frac{\alpha_m \eta_m}{\xi_0 (1 - \kappa)}
\Biggl[
\frac{4 \alpha_m \beta_m^2}{\xi_0} - \frac{x^2}{\alpha_m/\xi_0 + i x}
\Biggr],
\label{simple2}
\end{equation}
where $x = \xi/\xi_0$ and we used  $\chi_m-\chi_c = -4 \eta_m\alpha_m^2\beta_m^2$ and $\chi_m/\chi_c = 1 - \eta_m \beta_m^2$. 

For thin shells, we now can identify two regimes governed by the ratio of two small parameters,  $\eta_{m}=\rho^{(m)}\rho^{(s)}$ and $d/R_2$,
\begin{equation}
\frac{R_2}{d}\frac{\rho^{(m)}}{\rho^{(s)}} \approx \frac{M_m}{M_s},
\end{equation}
where $M_s$ is the metal shell mass, and $M_m$ is the mass of outer medium displaced by the core-shell particle. For a ``heavy shell'', $M_s\gg M_m$, the complex eigenvalue is given by
\begin{eqnarray}
\xi \simeq
\xi_0 - \frac{\lambda}{2}
\biggl[ \frac{\alpha_m-i\xi_0}{(\alpha_m/\xi_0)^2 + 1}- 4 \alpha_m\beta_m^2
\biggr],
\label{xi-heavy}
\end{eqnarray}
where $\xi_0$ is the eigenvalue for a nanoshell in vacuum and $\lambda= \alpha_m \eta_m/\xi_0 (1 - \kappa)$. In a good approximation, the real part is simply $\xi'\simeq \xi_0$, and is thus independent of the medium or aspect ratio. In contrast, the imaginary part, however small ($\xi'' \ll \xi'$), is only non-zero in the presence of outside medium and thus depends on both. Putting all together, we obtain in the ``heavy shell'' regime 
\begin{eqnarray}
\label{heavy}
&&
\omega^{(0)}
\simeq
\frac{ 2 c_L^{(s)}\beta_s}{R_2} \sqrt{3 - 4\beta_s^2},
\nonumber\\
&&
\gamma^{(0)}
\simeq
\frac{c_L^{(m)}}{d}
\frac{2\eta_m \beta_s^2 (3-4\beta_s^2)}
{\alpha_m^2 + 4\beta_s^2(3-4\beta_s^2)}.
\end{eqnarray}
Here the damping rate is determined by the shell thickness rather than by the overall size. In the opposite case of a ``light shell'', $M_s\ll M_m$, the eigenvalue is given by $\xi\simeq 2\alpha_m\beta_m\bigl(\sqrt{1-\beta_m^2} +i \beta_m\bigr)$, yielding
\begin{equation}
\label{light}
\omega^{(0)} \simeq 2c_T^{(m)}/R_2, ~~ \gamma^{(0)} \simeq \omega c_T^{(m)}/c_L^{(m)}. 
\end{equation}
Note that, in this case, if the transverse velocity of outside medium is zero (e.g., in water), both frequency and damping rate vanish. 

\begin{table}
\caption{\label{tab:parameters} Longitudinal velocity  $c_{L}$(m/s), transverse velocity $ c_{T} $(m/s), density $ \rho ($kg/m$^{3}$) acoustic impedance $ Z $(10$^{7}$kg/m$^{2}$s) and shear module $ \mu $(10$^{10}$kg/ms$^{2}$) of bulk gold, silver and  several dielectrics.}
\vskip 5mm
\begin{ruledtabular}
\begin{tabular}{llllll}
 &$c_{L}$ &$c_{T}$ & $\rho$ & Z  &  $\mu$\\
\hline
Au & 3240 &  1200 &  19700  & 6.38 & 2.84\\ 
Ag & 3650  & 1660  & 10400  & 3.80 & 2.87\\ 
Glass &  4610 &  2590  &  3810 & 1.76  &  2.56\\
Fused silica   &   5970  &   3765    &   2200   & 1.31  &   3.12 \\
Water &  1490 &  0 &  1000 & 0.15 & 0\\
\end{tabular}
\end{ruledtabular}
\end{table}
\begin{figure}[b]
\begin{center}
 \includegraphics[width=2.5in]{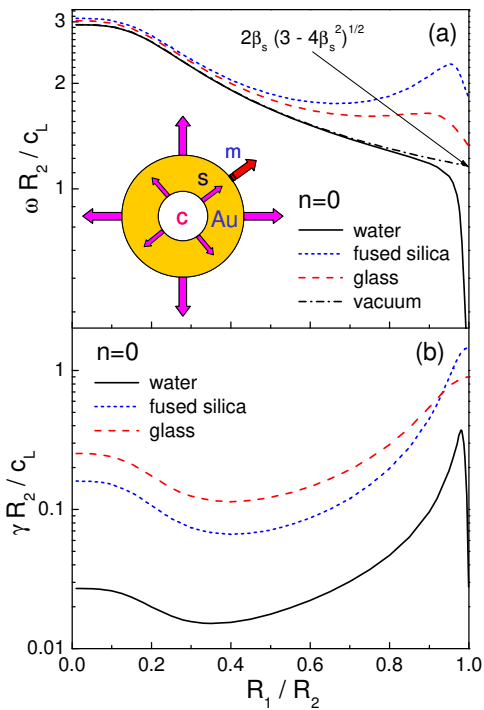}
\end{center}
\caption{\label{fig:nanoshell0} Normalized frequency (a) and damping rate (b) of  fundamental ($ n=0 $) radial mode in Au nanoshell vs. aspect ratio in various environments.  The cartoon illustrates the motion of shell boundaries.
}
\end{figure}

Below we present the results of numerical calculations of vibrational modes spectrum for Au nanoshells in three dielectric media: water, fused silica and glass (BaO-P$_{2}$O$_{5}$). The  material parameters are listed in Table \ref{tab:parameters}. In Fig. \ref{fig:nanoshell0} we show the energy and damping rate for fundamental ($n=0$) breathing mode versus aspect ratio $R_1/R_2$. In entire range of aspect ratios, the nanoshell frequencies are lower than those of solid particles of the same overall size. Both ``heavy" and ``light" regimes can be clearly distinguished. In the former regime, for aspect ratios $\kappa \leq 0.5$, the eigenfrequencies are only weakly dependent on media and follow each other, according to Eq.~(\ref{heavy}). For aspect ratios $\kappa \geq 0.5$, the eigenfrequencies start deviating from each other due to the effect of environment, and for very thin shell ($\kappa \gtrsim 0.9$) the ``light shell" behavior is observed, in agreement with Eq.~(\ref{light}). In contrast, the damping rate strongly depends on the environment for all $\kappa$, however, the two regimes are clearly seen. For thick shells, the damping is mostly determined by longitudinal sound velocities of outside media [see Eq.~(\ref{heavy})], while for thin nanoshells it is dominated by the transverse ones [see Eq.~(\ref{light}] so that the  curves for glass and fused silica cross each other. Importantly, in all cases, the damping rate exhibits a minimum at $\kappa\approx 0.4$, while decreasing shell thickness it behaves as $\gamma\propto d^{-1}$. Note that for thin nanoshell in water, both frequency and damping rate vanish due to the absence of transverse sound ($c_{T}^{water}=0$).
\begin{figure}[t]
\centering
\includegraphics[width=2.5in]{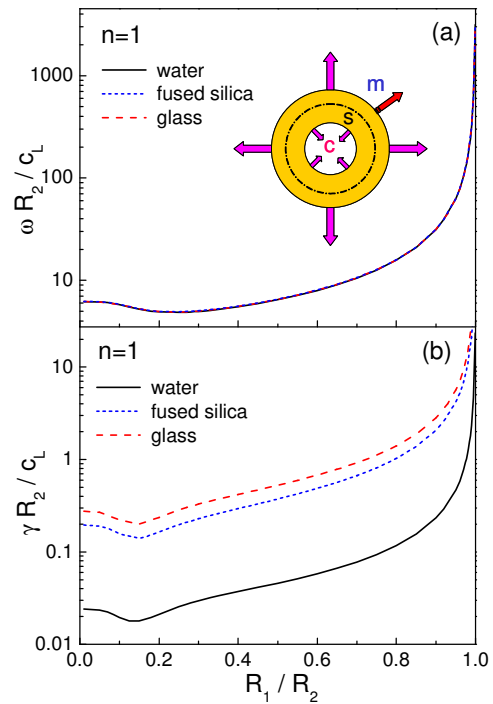}
\caption{
\label{fig:nanoshell1}
Normalized frequency (a) and damping rate (b) of $ n=1 $ radial mode in Au nanoshell.  Dashed line in cartoon indicates zero-displacement surface.
} 
\end{figure}

\begin{figure}[tb]
\centering
 \includegraphics[width=2.5in]{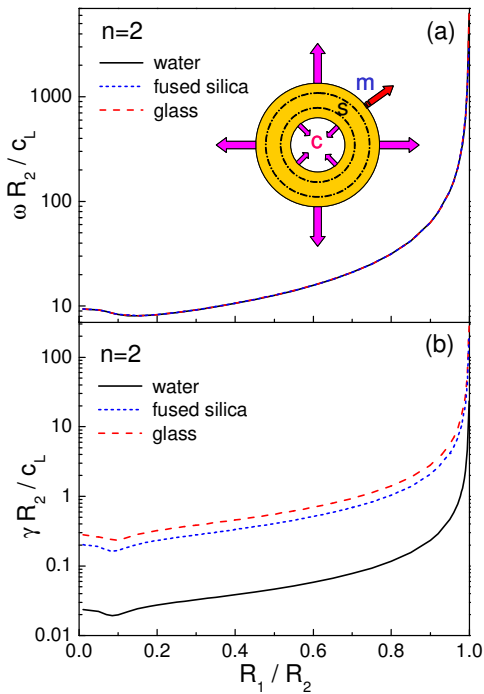}
\caption{\label{fig:nanoshell2}
Same as Fig.~\protect\ref{fig:nanoshell1}, but for $ n=2 $ mode.
 } 
\end{figure}

Situation is drastically different for higher modes, shown in Fig.~\ref{fig:nanoshell1} and \ref{fig:nanoshell2} for $ n=1 $ and $ n=2$, respectively. Both frequency and damping rate increase with aspect ratio and behave as $ d^{-1} $ for thin shells. The reason is that higher modes correspond to oscillations around a zero-displacement surface inside the shell rather than modulation of its overall size. In this case, the eigenfrequencies are nearly independent of environment in the entire range of aspect ratios, indicating the presence of only one (``heavy'' shell) regime. At the same time, the damping rate exhibits a weak minimum that shifts towards smaller $ \kappa $ as  $ n $ increases (compare Figs.~\ref{fig:nanoshell1} and \ref{fig:nanoshell2}).
\begin{figure}[tb]
\centering
 \includegraphics[width=2.5in]{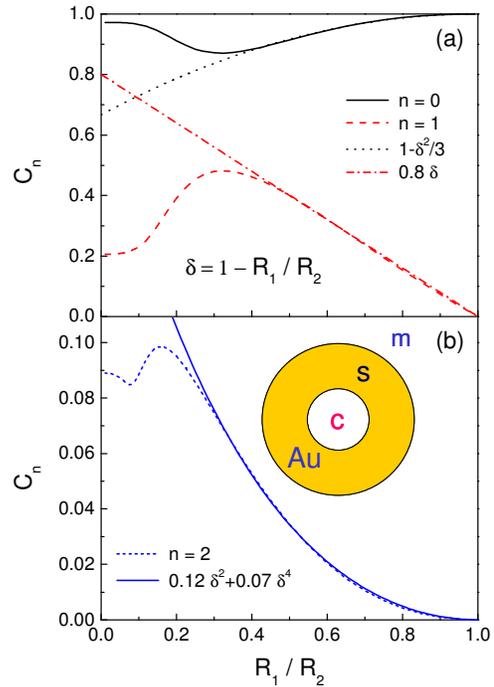}
\caption{\label{fig:shellosc} 
Oscillator strengths for $ n=0,1 $ (a) and $ n=2 $ (b) radial modes of a nanoshell in  vacuum vs. aspect ratio is shown together with analytical fit.
} 
\end{figure}

In Fig. \ref{fig:shellosc}, we show calculated oscillator strengths for $n=0, 1, 2$ modes versus aspect ratio for a nanoshell in vacuum. In solid nanoparticles, the oscillator strength for $n$-mode can be approximated by $ C_{n}^{(sol)} \approx 3\sqrt{10}/[\pi (n+1)]^{2} $. However these values slightly larger than those obtained numerically ($ \kappa =0$ in Fig.~\ref{fig:shellosc}). This discrepancy can be attributed to the corrections due to the second term in Eq. (\ref{final-xi}) (third term vanishes in vacuum) that reduce the oscillator strength of, e.g., $ n=1 $ mode from 0.25 to 0.2. In contrast, for nanoshells, $C_l$ vanishes in the thin shell limit $  $ for all higher modes $ l\neq 0 $.  For $\delta=d/R_{2}\ll 1$, the behavior of $C_n$ can be approximated as
\begin{eqnarray}
C_{0} \approx 1 - \frac{\delta^{2}}{3},~~
C_{1} \approx 0.8 \delta,~~
C_{2} \approx 0.12 \delta^{2}+0.07\delta^{4}.
\label{c0}
\end{eqnarray}
In the $d=0$ limit,  $C_0$ reaches it maximal value, $C_0=1$, i.e., the fundamental mode carries the entire oscillator strength. As a result, for \emph{thin} nanoshells, the fundamental mode is more pronounced as compared to solid particles.\cite{guillon-nl07} At the same time, for \emph{thick} nanoshells, $R_{1}/R_{2}   \approx 0.3$, the oscillator strength of $ n=1 $ mode is considerably higher than in solid particle (see Fig. \ref{fig:shellosc}).  This enhancement is less pronounced for higher modes.

\section{Vibrational modes of bimetallic  nanoparticle}
\label{sec:bimetallic}
Consider now bimetallic core-shell nanoparticles embedded in outside dielectric. As discussed in the previous section, the acoustical modes are determined by vibrations of both core and shell parts. Below we present results for eigenfrequencies and damping rates of Au/Ag core-shell spherical particles in different environments based on full numerical solution of Eqs.~(\ref{breathmodes}). Since complex eigenvalues $ \xi_{n} $ are normalized in terms of longitudinal sound velocity in the shell region, in order to obtain eigenvalues for solid Au sphere at $ \kappa=1 $ one simply needs to rescale  $\xi_{n} $ by $ c_{L}^{(Ag)}/c_{L}^{(Au)}$.

\begin{figure}[tb]
\centering
 \includegraphics[width=2.5in]{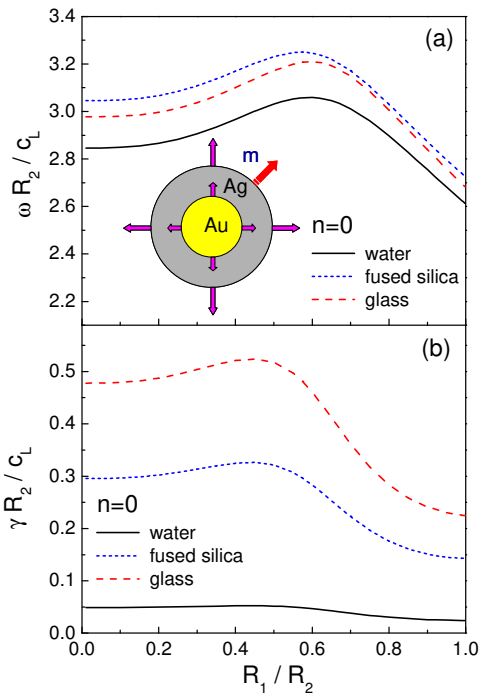}
\caption{\label{fig:coreshell0}
Frequency (a) and damping rate (b) of  $ n=0 $ radial mode in Au/Ag bimetallic particle, normalized to the shell  longitudinal sound velocity, in various environments.
} 
\end{figure}

In Fig.~\ref{fig:coreshell0}, we show the frequency and damping rate vs. aspect ratio in three different media. In contrast to nanoshells, here the frequency is higher than in either Ag and Au solid particles in a wide range of $ \kappa  $. Another distinction is a dependence of the breathing mode frequency on the environment in entire range of $ \kappa $. The sensitivity to the environment is much more pronounced for damping rate that shows a weak maximum for aspect ratios $ \kappa \approx 0.5 $.  In the course of transition from solid Ag to solid Au nanoparticles, the magnitude of damping drops roughly by factor of two reflecting the relation between acoustic impedances of Ag  and Au (see Table~\ref{tab:parameters}).
\begin{figure}[tb]
\centering
 \includegraphics[width=2.5in]{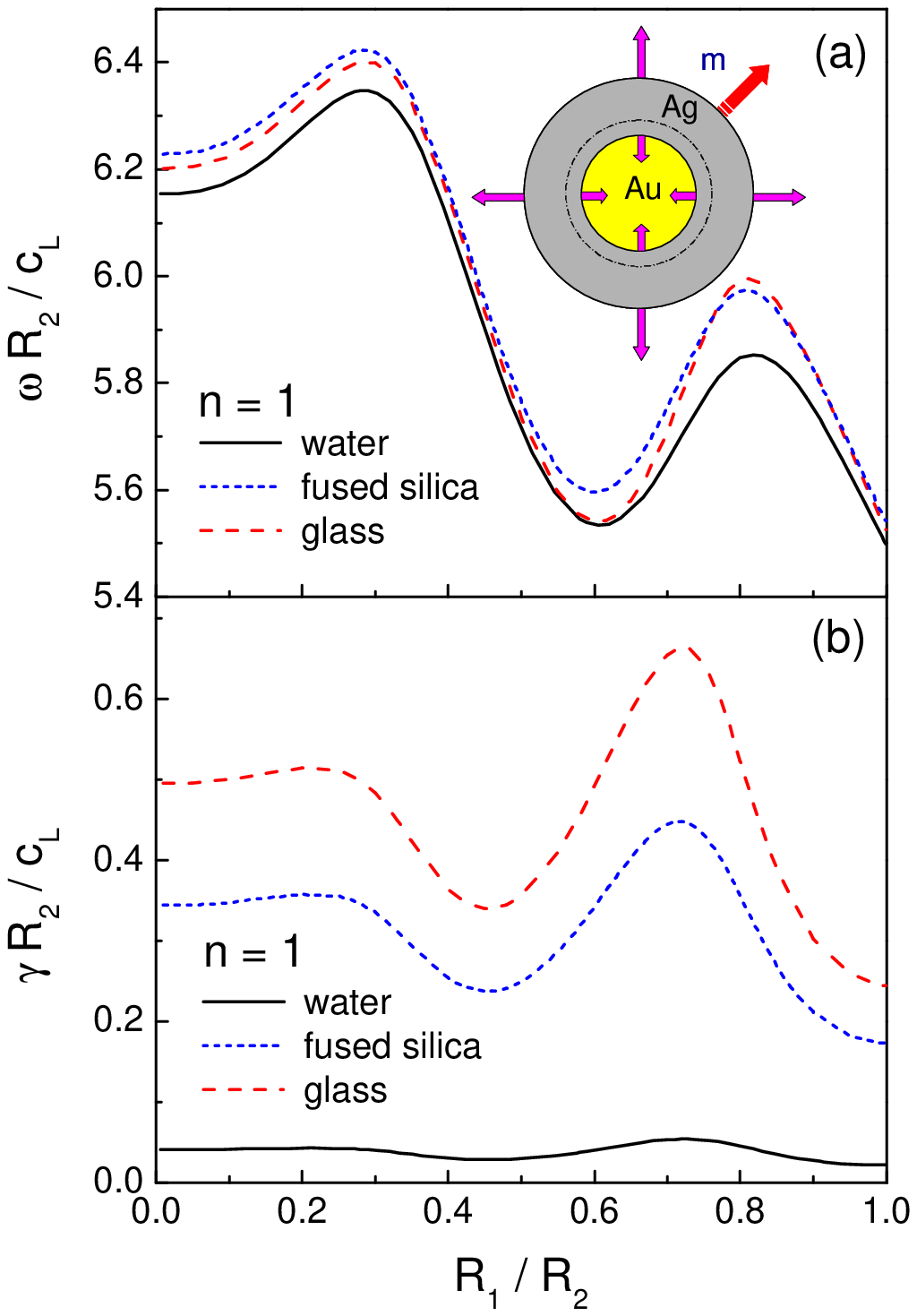}
\caption{\label{fig:coreshell1}
Same as Fig.~\protect\ref{fig:coreshell0}, but with $ n=1 $.} 
\end{figure}
\begin{figure}[tb]
\centering
 \includegraphics[width=2.4in]{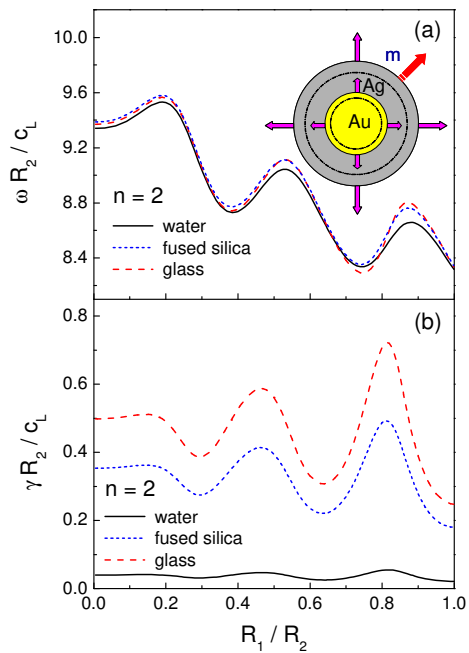}
\caption{\label{fig:coreshell2}
Same as Fig.~\protect\ref{fig:coreshell1}, but with $ n=2 $.
} 
\end{figure}

In Figs.~\ref{fig:coreshell1} and \ref{fig:coreshell2}, we show the result of calculations for $ n=1 $ and $ n=2 $ modes, respectively. The main feature distinguishing bimetallic particles from nanoshells is \emph{periodic dependence} of both frequency and damping rates on aspect ratio. The number of oscillations coincides with the mode number $ n $. Similar to nanoshells, the frequencies of higher modes only weakly depend on the environment, while the damping rate shows the opposite trend. Furthermore, the oscillation amplitude of damping rate increases with aspect ratio reaching about 50\%.  The origin of oscillations is the interference of cavity and shell modes with change of relative core and shell sizes.
\begin{figure}[tb]
\centering
 \includegraphics[width=2.5in]{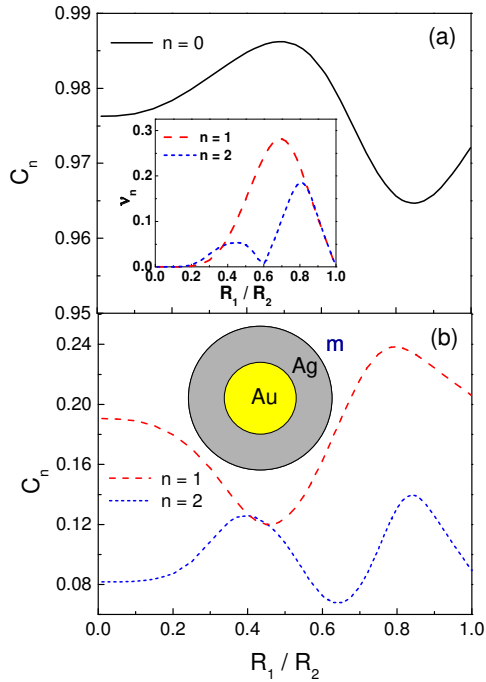}
\caption{\label{fig:coreshellosc}
Oscillator strengths are shown for $ n=0 $ (a) and $ n=1,2 $ (b) radial modes of Au/Ag particle in  vacuum vs. $ R_{1}/R_{2} $. Inset: admixture coefficients for $ n=1,2 $ modes. } 
\end{figure}

Oscillator strengths of $ n=0,1,2 $ modes of Au/Ag particle in vacuum, $ C_{n} $  are plotted in Fig.~\ref{fig:coreshellosc} vs. aspect ratio. As in solid particles and nanoshells, here the fundamental mode carries most of the oscillator strength, but slightly oscillates with aspect ratio. The oscillations are considerably stronger for $ C_{1} $ and $ C_{2} $, reaching about 50\% in peak-to-peak amplitude. Note that the higher modes, plotted  here, are the orthogonal states with amplitude $ v_{n} $ (see Eq.~\ref{newset1}) that carry an admixture from the fundamental mode, rather than the pure $ n=1 $ and $ n=2 $ states. The dependence of admixture coefficients $ \nu_{n} $ on $ \kappa $ is highly non-monotonic, reaching maximal value about 0.3 for $ n=1 $ mode and 0.2 for $ n=2 $ mode (see inset). This admixture is reducing the strength of $ n=1 $ mode, and enhancing that of $ n=2 $ mode, resulting in about the same oscillator strength for both modes for aspect ratio close to 0.4. Note, finally that, in contrast to nanoshells, higher modes should be more detectable for higher aspect ratios ($ \kappa \approx 0.8 $ for Au/Ag particle).

\section{Conclusions}
\label{sec:conclusions}
In summary, we studied the spectrum of radial vibrational modes in bimetallic core-shell nanoparticles and metal nanoshells in an environment. We found that the acoustical signature of these nanostructures is extremely sensitive to their composition, geometry and environment, and differs significantly from that of solid nanoparticles. In particular, for bimetallic nanoparticles, frequency and damping rate are periodic functions of the  aspect ratio, $ R_{1}/R_{2} $, with period determined by the mode number, while in nanoshells this dependence is monotonic. This difference between nanoshels and bimetallic particles can be traced to the fact that, in the latter, the core is fully engaged in vibrations, leading to the interference between cavity and shell modes. Another distinction is  much higher damping rate in thin nanoshells in comparison with solid or bimetallic particles. Thus, the acoustical probe of composite nanostructures can provide a sensitive tool for determination of their composition and geometry .

This work was supported in part by NSF under Grant No. DMR-0606509, by NIH under Grant No. 2 S06 GM008047-33, and by DoD under contract No. W912HZ-06-C-0057.


\begin{thebibliography}{}
%
%
\bibitem{halas-prl97}R. D. Averitt, D. Sarkar and N. J. Halas, 
Phys. Rev. Lett. {\textbf 78}, 4217 (1997) .

\bibitem{halas-nanorice} H. Wang, D. Brandl, F. Le, P. Nordlander and N. J. Halas,
Nano Lett. \textbf{6}, 827 (2006).

\bibitem{halas-matryoshka} E. Prodan, C. Radloff, N. J. Halas, and P. Nordlander, Science \textbf{302}, 419 (2003).


\bibitem{halas-prb02} S. L. Westcott, J. B. Jackson, C. Radloff,  N. J.  Halas, Phys. ReV.
B \textbf{66}, 155431 (2002).

\bibitem{prodan-nl02} E.  Prodan, P. Nordlander,  N.  J. Halas,  Nano Lett. \textbf{3}, 1411 (2002).

\bibitem{klar-nl04}G. Raschke, S. Brogl, A. S. Susha, A. L. Rogach,  T. A. Klar, J. Feldmann, B. Fieres, N. Petkov, T. Bein, A. Nichtl, K. Kurzinger, Nano Lett. \textbf{4}, 1853 (2004).

\bibitem{halas-nl04} Nehl, C. L.; Grady, N. K.; Goodrich, G. P.; Tam, F.; Halas, N. J.;
Hafner, J. H. Nano Lett. \textbf{4}, 2355 (2004).

\bibitem{klar-prl98}T. Klar, M. Perner, S. Grosse, G. von Plessen, W. Spirkl, J. Feldmann, Phys. ReV. Lett. \textbf{80}, 4249 (1998).

\bibitem{halas-ap01} S. R. Sershen,  S. L. Westcott, J. L. West, N. J. Halas,  Appl. Phys. B \textbf{73}, 379 (2001).

\bibitem{sun-ac02} Y. Sun, Y.  Xia,  Anal. Chem. \textbf{74},  5297 (2002).

\bibitem{halas-apl03} J. B. Jackson, S. L. Westcott, L.  R. Hirsch, J. L. West, N. J.  Halas, Appl. Phys. Lett. \textbf{82}, 257 (2003).

\bibitem{halas-ac03} L. R. Hirsch,  J. B. Jackson, A. Lee, N. J. Halas, J. L. West, Anal. Chem. \textbf{75}, 2377 (2003).

\bibitem{halas-nl02} C. Loo, A. Lowery, N. J. Halas, J. West, R. Drezek, Nano Lett. \textbf{5}, 709
(2002).


\bibitem{west-pnas03} L. R. Hirsch, R. J. Stafford, J. A. Bankson, S. R. Sershen, B. Rivera, R. E. Price, J. D. Hazle, N. J. Halas, and J. L. West, 
PNAS \textbf{100}, 13549 (2003).

\bibitem{vallee-jcp99} N. Del Fatti, C. Voisin, F. Chevy, F. Vall\'{e}e, and C. Flytzanis,  J. Chem. Phys. {\bf 110},  11484 (1999).


\bibitem{hartland-jcp99} J. S. Hodak, A. Henglein, and G. V. Hartland,
J. Chem. Phys. {\bf 111},  8613 (1999).

\bibitem{vallee-jcp01} H. Portales, L. Saviot, E. Duval, M. Fujii,
  S. Hayashi, N. Del Fatti, and F. Vall\'{e}e, 
J. Chem. Phys. {\bf 115}, 3444 (2001).

\bibitem{elsayed-nl04}
W. Huang, W. Qian, and M. A. El-Sayed, Nano Lett. {\bf 4}, 1741 (2004).

\bibitem{orrit-prl05}
M. A. van Dijk, M. Lippitz, and M. Orrit, Phys. Rev. Lett. {\bf 95}, 267406 (2005).

\bibitem{hartland-bimetal1} J. H. Hodak, A. Henglein, and G. V. Hartland, J. Phys. Chem. B 
{\bf 104},  5053 (2000).

\bibitem{hartland-bimetal2} J. E. Sader, G. V. Hartland, and P. Mulvaney, J. Phys. Chem. B {\bf 106},  1399 (2000).

\bibitem{hartland-nl07} 
H. Petrova, C.-H. Lin, M. Hu, J. Chen,  A. R. Siekkinen, Y. Xia, J. E. Sader,  G.V. Hartland,
Nano Lett. {\bf 7}, 1059 (2007). 

\bibitem{guillon-nl07} C. Guillon, P. Langot, N. Del Fatti, F. Vall\'{e}e, A. S. Kirakosyan, T. V. Shahbazyan,T. Cardinal and M. Treguer,
Nano Lett. {\bf 7}, 138 (2007).


\bibitem{vallee-jpcb01} C. Voisin, N. Del Fatti, D. Christofilos, and F. Vall\'{e}e,  J. Phys. Chem. B {\bf 105},  2264 (2001)  and references therein.

\bibitem{love-elast} A. E. H. Love, {\it A Treatise on Mathematical Theory of
    Elasticity}, Dover, New York (1944).

\bibitem{dubrovskiy81} V. A. Dubrovskiy and V. S. Morochnik,  Izv. Earth Phys. {\bf 17},  494 (1981).


\bibitem{vallee-pb02} C. Voisin, D. Christofilos, N. Del Fatti, F. Vall\'{e}e, 
Physica B {\textbf 89}, 316 (2002). 

\bibitem{vallee-ass04}	A. Neleta, A. Cruta, A. Arboueta, N. Del Fatti, F. Vall\'{e}e, H. Portales, L. Saviotc, E. Duval,
Appl. Surf. Sci. \textbf{226}, 209 (2004).

\end{thebibliography}
\end{document}